**Title:** When Technology Isn't Enough: Insights from a Pilot Cybersecurity Culture Assessment in a Safety-Critical Industrial Organisation

**Authors:** Tita Alissa Bach[1], Linn Pedersen[1], Maria Kinck Borén[2], Lisa Christoffersen Temte[2]

Contact information for readers: Tita Alissa Bach, Tita.Alissa.Bach@dnv.com, +4791124056

Word count: 10585

[1] Group Research and Development, DNV, 1363 Høvik, Norway

[2] Safety Risk Nordic, DNV, 1363 Høvik, Norway

**Abstract**:

As cyber threats increasingly exploit human behaviour, technical controls alone cannot ensure organisational cybersecurity (CS). Strengthening cybersecurity culture (CSC) is especially vital in safety-critical industries, yet empirical research on CSC in real-world industrial settings remains scarce. This paper addresses this gap through a pilot mixed-methods CSC assessment in a global safety-critical organisation.

The study examined employees' CS knowledge, attitudes, behaviours, and the organisational and contextual factors influencing them. We conducted a survey and semi-structured interviews with employees at DNV, a global organisation working in safety-critical industries. We selected two countries based on internally administered phishing simulations: Country 1 had generally stronger performance, Country 2 weaker. This contrast provided insight into CSC dynamics and guided targeted strategies. In Country 1, 258 employees were invited (67% response); in Country 2, 113 were invited (30%). Interviews included 20 participants from Country 1 and 10 from Country 2.

Overall CSC profiles were similar, yet each showed distinct enablers and challenges. Both demonstrated strong phishing awareness and saw CS as a business priority. However, most perceived phishing as the main risk and lacked clarity on handling other incidents. Few knew the content or location of CS policies. Line managers were default contacts, though follow-up on reported concerns was often unclear. Participants stressed aligning CS expectations with job relevance and workflows.

Underlying contributors to differences in CSC profiles: Country 1 had many external employees, who did not always receive the organisation's official PCs and had limited access to CS training and policies, highlighting an opportunity to strengthen monitoring consistency. In Country 2, the low survey response likely stemmed from their "no-link in emails" policy; some mistook the survey invitation as phishing. This policy may have contributed to stronger phishing performance by fostering caution but also highlighted inconsistencies in CS practices.

Our findings show that fostering a resilient CSC requires consistent leadership involvement, targeted communication, tailored improvement measures, alignment between CS policies and practices, and regular CSC assessments. Embedding these into strategy will complement technical defences and enhance sustainable CS in safety-critical settings.

# Introduction

In March 2019, Norsk Hydro, one of the world's largest aluminium and energy companies, fell victim to a major ransomware attack that impacted approximately 35,000 employees across 40 countries. The breach occurred when an employee unknowingly opened a malicious email attachment from what appeared to be a trusted customer. The ransomware, identified as LockerGoga, infiltrated the company's IT systems, encrypted files on laptops and servers and displayed ransom notes on the screens of the corrupted computers demanding payment in Bitcoin. Rather than capitulating to the attackers' demands, Norsk Hydro chose to be transparent about the incident, declined to pay the ransom, and mobilised CS experts to restore operations. The attack caused production losses at multiple plants and high recovery costs that were estimated to total $71 million and damaged the reputation of the company [1-3].

In addition to financial and reputational damage to organisations, cyberattacks can pose serious risks to public safety. For example, in February 2021, unknown cyber actors gained unauthorized access to the Supervisory Control and Data Acquisition (SCADA) system of a US water treatment facility and attempted to increase the dosage of a hazardous chemical [4], potentially leading to public health crisis. Plant personnel quickly detected and corrected the change, preventing any harm. The breach likely exploited the risks of using unsupported software and weak CS practices: poor password security, shared passwords among multiple operators, outdated operating systems, and vulnerable desktop sharing software.

Cyber vulnerabilities can also affect critical infrastructure. A ransomware attack in May 2021 on the Colonial Pipeline, a major US fuel artery, caused significant disruptions to fuel distributions along the eastern seaboard [5, 6]. The attackers gained access to the system using a compromised password for an inactive Virtual Private Network (VPN) account that did not have multi-factor authentication enabled. The ransomware, identified as DarkSide, targeted Colonial's IT systems and led Colonial Pipeline to shut down the pipeline as a precautionary measure to prevent further compromise. It was reported that Colonial Pipeline paid a ransom of $4.4 million in cryptocurrency to regain access to their data and systems [5, 6].

These attacks are primarily the result of scale and human factors. *Microsoft Digital Defense Report* covering trends from July 2023 through June 2024 shows that their customers globally are facing 600 million attacks daily. Verizon [7], the world's second-largest telecommunications company by revenue, revealed in its 2024 report that 68% of data breaches in 2024 were caused by human error, including social engineering scams [8]. A 2024 IBM report shows that the average cost of a data breach reached a record high of $4.88 million, a 10% increase from the previous year [9]. This same report found that human error and IT failures account for nearly half of all CS breaches, demonstrating that human factors contribute as much to these incidents as vulnerable technological systems [9]. A 2024 ProofPoint survey of 8550 users and security professionals across 15 countries found that 71% of users admitted to taking risky actions, 96% of whom were aware of the risk [10]. The top five reported risky behaviours by users included using work devices for personal

activities, reusing or sharing passwords, and connecting to public Wi-Fi without a VPN, responding to a message (email or SMS text) from someone they did not know, and accessing inappropriate websites.

Technology alone is insufficient to prevent these breaches. As cyber threats increasingly exploit human behaviour [11], organisations must foster more resilient CSC to effectively manage human-related vulnerabilities [12]. Reegård and Blackett [13] suggest that CSC should be defined as "the cybersecurity culture of the organisation that will influence the deployment and effectiveness of the cybersecurity management resources, policies, practices and procedures as they represent the work environment and underlying perceptions, attitudes, and habitual practices of employees at all levels" [13, p. 4041].

Although the concept of CSC has gained growing attention in academic literature, a critical gap remains in understanding how it is manifested and maintained in real-world industrial settings, particularly within safety-critical industries and critical infrastructure. In such environments, as the examples above show, the impact of a successful cyberattack can go far beyond financial or reputational damage to pose risks to human lives and public safety [14]. To date, much of the research has centred on individual awareness and technical measures, with limited empirical exploration of how CSC functions at the organisational level in operational contexts.

This paper addresses that gap by investigating CSC within an industrial setting operating on safety-critical industries and critical infrastructure through a pilot mixed-methods study. Conducted by an industry actor across selected business units in two countries, our study aims to provide insights on how CSC can be assessed and understood in practice. We combine quantitative survey data with qualitative insights from interviews to explore both attitudinal and contextual dimensions of CSC [15, 16].

We assessed employees' CS knowledge, attitudes, and behaviours; identified organisational and systemic factors that shape CSC; and explored how cultural and structural differences across regions may influence CS practices. By doing so, we aimed to identify barriers to behavioural change that are often overlooked in technology-centric approaches.

We pose two research questions: (1) How can CSC be meaningfully assessed in industrial settings using a mixed-methods approach? and (2) What types of cultural and organisational patterns emerge from a pilot assessment of CSC in a real-world industrial context?

By situating this research within an industry-led initiative, we contribute to both the academic discourse on CSC and to practical guidance for organisations. Our findings offer data-driven insights that can inform tailored CS strategies and highlight the value of integrating workplace-based cultural assessments into broader cyber resilience efforts.

## Methodology

We used a sequential mixed methods approach this study, conducting a survey prior to interviews [15-19]. Surveys and semi-structured interviews are complementary ways of understanding the complex nature of the human behaviours and attitudes behind building

CSC [20]. Surveys are a particularly suitable method for benchmarking and comparative analysis because they describe behaviours. However, while surveys are effective at identifying *what* is happening, they often fall short in explaining *why* these patterns occur. To gain deeper insight into the findings, we integrated semi-structured interviews into the research design. A key advantage of this interview format is that it allows participants to elaborate on their experiences and share subjective insights on topics they find relevant while the interviewer can still guide the conversation in a purposeful direction [21].

## The pilot setting

This pilot study was part of an internal pilot assessment of CSC within Det Norske Veritas (DNV) [22], a global assurance and risk management company headquartered in Norway. As DNV operates in over 100 countries across safety-critical sectors such as maritime, energy, and healthcare, understanding and strengthening CSC is essential to its broader risk management and digital resilience efforts. The study was initiated as a separate project within the internal Cybersecurity Game Changer programme. This programme aims to build lasting capabilities across DNV to reduce CS risks, strengthen resilience, and increase organisational CS maturity. It focuses on enabling change and delivering capabilities across four strategic areas to support progress along the CS maturity ladder.

Given the complexity of CSC and the diversity across DNV's global operations, we conducted a pilot study to test the methodology, refine the instruments, and assess practical implementation before scaling [23]. The pilot allowed for deeper understanding of contextual factors, ensured engagement from selected units, and helped avoid survey fatigue. It also identified potential barriers and areas of improvement to inform a more effective full-scale rollout.

We selected the two countries for our pilot study based on internal phishing simulation test results, an organisational effort to strengthen employees' ability to recognize phishing attempts through regularly deployed, varying-difficulty simulations. Country 1 had generally performed better than other countries, while Country 2 had generally performed worse. The reasons behind this performance gap were unclear, making these two countries strong candidates for exploring whether differences in CSC could help explain the results.

We invited 258 internal and external employees from one business area in Country 1 and 113 internal and external employees from several units in another business area in Country 2. External employees in our study are individuals who were not directly employed by DNV but worked on DNV-related projects or within DNV offices through third-party contracts, consultancies, or outsourcing arrangements. We included only those external employees who had a DNV email address, as this ensured they had regular access to internal communication channels and systems and were integrated into the digital workflows relevant for assessing CSC. All participating units operated in one or more safety-critical industries and/or critical infrastructure. We selected units and business areas in coordination with the Country Chairs, people who represent the organisation within specific countries and are responsible for coordination and cooperation across business areas. The decisions were based on practical considerations and the potential for mutual benefit between the pilot and selected units. For example, we gave priority to units that typically receive fewer surveys to avoid survey fatigue and to units or business areas where managers had expressed interest in participating. Additionally, we included units we considered particularly relevant for exploring CS tendencies.

### Ethical considerations

The data collection, analysis, and reporting methods ensured anonymity, confidentiality, and voluntary participation. The survey was anonymous, and we only used aggregated data in analysis and reporting. Participation in both the survey and interviews was entirely voluntary. All interviewees gave informed consent and were made aware of their right to refuse or withdraw from the study at any time without giving a reason. Interview recordings were accessible only to the five interviewers and the scribes. We anonymised all interview notes and transcripts and removed identifiable information prior to analysis. The data was stored securely on restricted systems in compliance with institutional data protection guidelines. Participants were informed that the findings would be used solely for the project's purposes and that their responses would not be linked to them personally or to their units in any form of evaluation.

### Stage 1: Survey

#### Development of the questionnaire

The short questionnaire was designed to minimize the time employees spent away from billable work and to encourage higher response rates. We utilised selected dimensions from two validated questionnaires: The Human Aspects of Information Security Questionnaire (HAIS-Q) [24], and the Cybersecurity Scale (CS-S) [25]. Additionally, we developed a "Governance" dimension to measure familiarity with the organisation's CS related policies, procedures, and guidelines, as this topic was not covered in HAIS-Q or CS-S. See Appendix 2 and Table 1 for all the questionnaire items used in our survey. The questionnaire responses were scored on a 5-point Likert scale where 1 stood for "Strongly Disagree", 2 for "Disagree", 3 for "Neutral", 4 for "Agree", and 5 for "Strongly Agree".

The questionnaire included eight dimensions and one independent item. It comprises 12 demographic questions and 62 items measuring knowledge, attitudes, and behaviours related to CS (see Appendix 1, Appendix 2, and Table 2). Although CS-S does not explicitly categorize its items into knowledge, attitudes, and behaviours like HAIS-Q does, we applied our judgment to map the CS-S Integrity items to these categories following the same approach used for our own dimension "Governance" and independent item "Priority". We iterated the questionnaire based on feedback from the project team and several other colleagues for clarity and fit-for-purpose. The final version was approved by the project team and the programme manager. Below, we discuss each of the validated questionnaires.

Table 1. The survey eight dimensions and their short description (N=62)

| Questionnaire | Dimensions | N | Description / examples |
|---|---|---|---|
| HAIS-Q | Password Management | 9 | Knowledge, attitudes, and behaviours related to using the same passwords, sharing passwords, using a strong password |
| HAIS-Q | Email Use | 9 | Knowledge, attitudes, and behaviours related to clicking on links and opening attachments in emails from known or unknown senders |
| HAIS-Q | Internet Use | 9 | Knowledge, attitudes, and behaviours related to downloading files, accessing dubious websites, entering information online |
| HAIS-Q | Social Media Use | 9 | Knowledge, attitudes, and behaviours related to social media privacy settings, consequences considerations, posting about work |
| HAIS-Q | Mobile Devices | 9 | Knowledge, attitudes, and behaviours related to physically securing mobile devices, sending sensitive information via Wi-Fi, shoulder surfing |
| HAIS-Q | Incident Reporting | 9 | Knowledge, attitudes, and behaviours related to reporting suspicious behaviours, ignoring poor security behaviours by colleagues, reporting all incidents |
| CS-S | Integrity | 4 | Knowledge, attitudes, and behaviours related to providing quality of control or ownership in the cyberspace |
| Own | Governance | 3 | Knowledge, attitudes, and behaviours related to the organisation's CS policies, procedures, and guidelines |
| Own | Priority | 1 | "My business has prioritised raising CS awareness in the past 12 months" |

### *The Human Aspects of Information Security Questionnaire*

The HAIS-Q instrument was used to assess individual-level CS knowledge, attitudes, and self-reported behaviours across several key domains. In this study, we omitted one of the original seven dimensions (i.e. Information Handling) to reduce survey length and participant burden. This decision was based on both practical considerations and an internal assessment of the dimension's relevance in the context of our organisational setting. Specifically, several items within the Information Handling dimension (e.g. practices related to physical printouts and USB device use) were less pertinent due to the digital-first and highly regulated nature of work practices within the organisation, where physical information handling is minimal or already tightly controlled through technical safeguards and policy enforcement.

The remaining six dimensions of the HAIS-Q cover core areas of CSC (i.e. Password Management, Email Use, Internet Use, Mobile Computing, Incident Reporting, and Social Media), and are well-aligned with the key risk areas identified in the organisation's CS strategy. One item in the Social Media dimension was rephrased for clarity from "It doesn't matter if I post things on social media that I wouldn't normally say in public" to "I feel comfortable posting on social media things I wouldn't normally say in a public setting".

### *The Cybersecurity Scale*

The CS-S was developed to assess individuals' CS related practices and perceptions across six dimensions: confidentiality, control/possession, integrity, authenticity, availability, and utility. We excluded all but the Integrity dimension to reduce survey length and avoid conceptual overlap with the HAIS-Q which already captures key aspects of individual CS behaviour and attitudes. The Integrity dimension was selected because it addresses elements not explicitly emphasised in the HAIS-Q, namely individuals' adherence to principles of honesty, consistency, and trustworthiness in CS practices. We slightly adapted the questionnaire items for clarity and contextual relevance by replacing the term "cyberspace" with "digital platforms," a term more commonly understood and aligned with the language used within the organisation (see Table 1 and Appendix 2). This minor modification preserved the original meaning while enhancing comprehensibility for participants.

### *Own "Governance" and "Priority"*

To complement the two validated questionnaires used in this study, we developed and incorporated an additional dimension labelled "Governance" to capture aspects not fully addressed by the existing instruments. The Governance dimension focused on employees' knowledge, attitudes, and behaviours towards the organisation's CS policies, procedures, and guidelines. This addition explicitly measured participants' engagement with formal governance documents and reflected the critical role that clear policy knowledge and compliance play in fostering an effective CSC.

In addition, we introduced a single independent item to assess how participants perceived their business areas prioritised CS within the past twelve months. This item captures employees' perceptions of organisational commitment and leadership focus on CS awareness and serves as an indicator of the contextual environment that supports CSC development. Including this item offered a valuable perspective on how strategic emphasis on CS awareness varied across business areas and how it might influence individual behaviours and attitudes toward CS practices.

### *The demographic questions*

Eight demographic questions identified differences in CSC responses across key employee variables (Appendix 1 and Table 2). Variables such as country, work context, employment type, and managerial role helped detect patterns and tailor recommendations to specific groups. For example, work context (e.g. field vs. office-based roles) could influence CS behaviours due to differences in daily tasks, access to systems, and exposure to cyber threats. Similarly, whether an employee was recruited through a merger and acquisition (M&A) process might also affect CSC, as those not originally part of the organisation might have received different onboarding, training, or hold alternative norms and practices related to CS.

### Survey participants and data collection

We invited all employees within the pilot setting to participate in the survey: 258 employees from one business area in Country 1 and 113 employees from several units in one business area in Country 2. We sent an email invitation with a link to the survey created on Microsoft Forms to all participants, copying the relevant Country Chairs for encouragement. The survey was open in Country 1 from 13 January to 31 January 2025, and in Country 2 from 20 January through to 5 February 2025. Several reminders were sent by emails copying the Country Chairs.

### Survey data analysis

To ensure consistency in scoring direction, negatively worded items were reverse scored so that higher survey scores were always interpreted as more positive than lower scores. Response categories were then numerically coded on a five-point Likert scale, with 1 representing the least favourable response (e.g. *strongly disagree*) and 5 the most favourable (e.g. *strongly agree*). We then computed mean scores for each of the eight dimensions, which served as the dependent variables in the analysis. The survey responses were analysed using the Jeffreys's Amazing Statistics Programme (JASP) [26]. Normality tests (Shapiro-Wilk) indicated that the data did not meet the assumptions for parametric analysis, so we applied non-parametric statistical tests to examine differences between and within groups across a set of independent variables (Table 2). After comparing the dependent variables between countries, we conducted the survey analyses within country using group comparisons. These assessed how CS knowledge, attitudes, and behaviours differed by demographic, contextual, and organisational factors within each country. Where applicable, Mann-Whitney $U$ tests and Kruskal-Wallis tests identified statistically significant differences in mean scores across the identified dimensions. In addition, open ended survey answers and interview responses that included general suggestions for improving the organisation's CS were combined and analysed qualitatively. Similar responses were grouped into themes, organised by country.

Table 2. The CSC's survey's independent and dependent variables

| Independent variables | Dependent variables |
|---|---|
| - Country<br>    o Country 1<br>    o Country 2<br>- Employment type:<br>    o Permanent<br>    o Temporary<br>    o External<br>- Gender<br>    o Male<br>    o Female<br>    o Prefer not to say<br>- Age group<br>    o >25 years old<br>    o 25-30 years old | Nine CSC's dimensions:<br>1. Password Management<br>2. Email Use<br>3. Internet Use<br>4. Social Media Use<br>5. Mobile Devices<br>6. Incident Reporting<br>7. Integrity<br>8. Governance<br>9. CS as a priority |

- o 31-40 years old
  - o 41-50 years old
  - o >50 years old
  - o Prefer not to say
- Tenure at the organisation
  - o <1 year
  - o 1-5 years
  - o 6-10 years
  - o 11-20 years
  - o 21-30 years
  - o >30 years
- Work context
  - o Field employees (e.g. surveyors, assessors, auditors)
  - o Office-based employees
- Managerial role
  - o Line managers
  - o Non-managers
- Nature of recruitment
  - o Recruited directly by the organisation (not through an M&A process)
  - o Not recruited directly by the organisation (e.g. through an M&A process)

### Stage 2: Semi-structured interviews

We used semi-structured interviews get a deeper understanding of CSC and the survey responses. This style of interview was more conversational and made sure all interview participants were asked the same questions while allowing flexibility to personalize the interview to each participant. The project team, programme manager, and a data management and governance manager all contributed to developing the interview questions (see Appendix 3). The questions were tested in two mock-up interviews within the project team with one interviewer and one scribe. Based on the feedback and mock-up evaluation, the project team revised the questions iteratively and approved the final set of questions. The team also discussed and agreed on an interview procedure to ensure consistency across interviewers and scribes. There were four main interviewers and scribes who had backgrounds in human factors and psychology and extensive interview experience.

#### Interview participants and data collection

We used random sampling to invite participants and scheduled 30–45-minute interviews between1 February to 28 February 2025. We invited the participants via Microsoft Teams so interviewers and scribes could access potential participants' the availability. Invited participants could accept, decline, or propose new time. Interviews usually included an interviewer and scribe plus Microsoft Co-pilot to help summarize for validation. Due to limited resources, sometimes the interviewer was also the scribe. We obtained informed consent from participants before recording. Only one interview participant refused to be recorded.

### Interview qualitative data analysis

We conducted a thematic analysis of the interview data following an inductive approach. We identified and extracted significant words and phrases from each transcript and interpreted their meanings in context. Codes expressing similar meanings were grouped into themes. Both common and divergent viewpoints were identified and highlighted. We also noted patterns that aligned or contrasted with the survey findings. The four interviewers and scribes collaboratively coded responses during iterative meetings and disagreements were discussed and resolved by consensus. We then defined and organised themes and sub-themes based on the structure of the interview guide. This process aimed to ensure a rigorous and transparent analysis of the qualitative data. In addition, we combined and qualitatively analysed interview responses and open-ended survey answers that included general suggestions for improving the organisation's CS. Similar responses were grouped into themes and organised by country.

## Results

### Stage 1: Survey results

We sent the questionnaire to 258 (Country 1) and 113 (Country 2) employees and received a 67% response rate (Country 1) and a 30% response rate (Country 2). The demographic characteristics of the survey participants are listed in Table 3 for each country.

Table 3. Survey participants' demographic characteristics by country

| Demographic characteristics | | Country 1 | | Country 2 | |
|---|---|---|---|---|---|
| | | Frequency | Percentage | Frequency | Percentage |
| Employment type | Permanent | 73 | 41.95% | 33 | 97.06 % |
| | Temporary | 4 | 2.30 % | 1 | 2.94 % |
| | External | 97 | 55.75 % | 0 | 0 |
| | Total | **174** | **100%** | **34** | **100%** |
| Gender | Male | 142 | 81.61 % | 24 | 70.59 % |
| | Female | 30 | 17.24 % | 10 | 29.41 % |
| | Prefer not to say | 2 | 1.15 % | 0 | 0 |
| | Total | **174** | **100%** | **34** | **100%** |
| Age group | >25 years old | 0 | 0 | 0 | 0 |
| | 25-30 years old | 5 | 2.87 % | 4 | 11.76 % |
| | 31-40 years old | 25 | 14.37 % | 10 | 29.41 % |
| | 41-50 years old | 39 | 22.41 % | 11 | 32.35 % |
| | >50 years old | 102 | 58.62 % | 8 | 23.53 % |
| | Prefer not to say | 3 | 1.72 % | 1 | 2.94 % |
| | Total | **174** | **100%** | **34** | **100%** |
| Tenure at the organisation | <1 year | 13 | 7.47 % | 1 | 2.94 % |
| | 1-5 years | 57 | 32.76 % | 7 | 20.59 % |
| | 6-10 years | 39 | 22.41 % | 7 | 20.59 % |
| | 11-20 years | 56 | 32.18 % | 12 | 35.29 % |
| | 21-30 years | 9 | 5.17 % | 5 | 14.71 % |
| | >30 years | 0 | 0 | 2 | 5.88 % |
| | Total | **174** | **100%** | **34** | **100%** |
| Work context | Field employees (e.g. surveyors, assessors, auditors) | 134 | 77.01 % | 0 | 0 |
| | Office-based employees | 40 | 22.99 % | 34 | 100.00 % |
| | Total | **174** | **100%** | **34** | **100%** |
| Managerial role | Line managers | 16 | 9.20 % | 12 | 35.29 % |
|  | Non-managers | 158 | 90.80 % | 22 | 64.71 % |
|  | **Total** | **174** | **100%** | **34** | **100%** |
| Nature of recruitment | Recruited directly by the organisation (not through an M&A acquisition process) | 154 | 88.51 % | 23 | 67.65 % |
|  | Recruited through a merger and acquisition process | 15 | 8.62 % | 11 | 32.35 % |
|  | Do not know | 5 | 2.87 % | 34 | 100% |
|  | **Total** | **174** | **100%** | **23** | **67.65 %** |

The most common primary roles were auditors (Country 1) and consultants (Country 2) (Figure 1 and Figure 2). The most used electronic devices on a daily basis for both countries were the office PCs followed by mobile phones (Figure 3 and Figure 4). Country 1 had some survey participants who also used personal PCs for work on a daily basis, whereas Country 2 had none. The general improvement suggestions collected via the open-ended survey question and the interviews were categorised by themes and reported by country (Appendix 4).

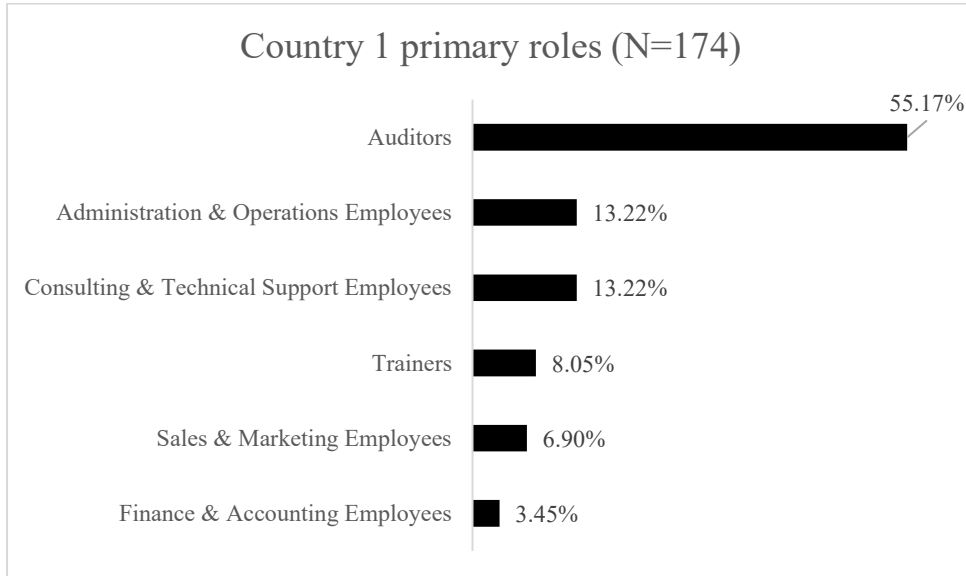

Figure 1. Primary roles of survey participants in Country 1

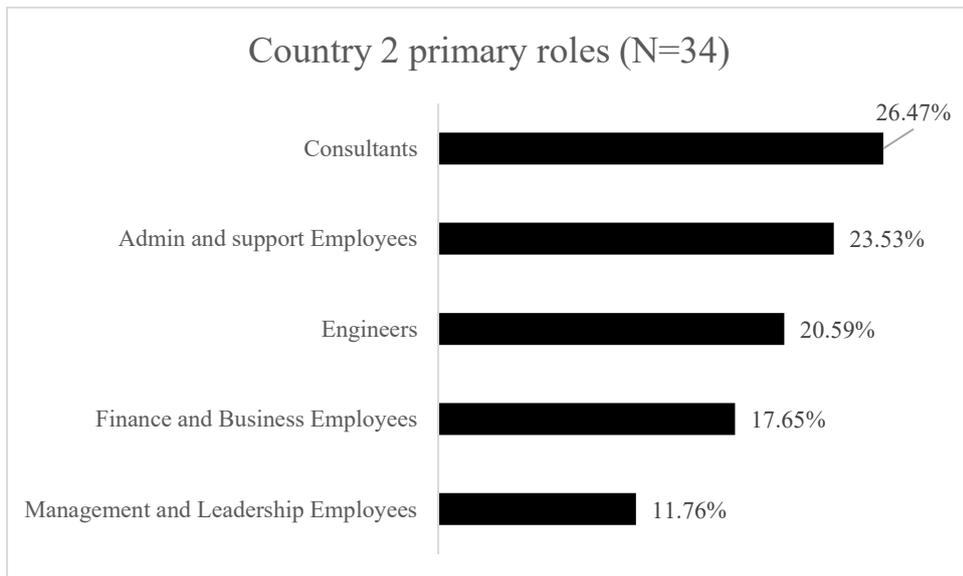

Figure 2. Primary roles of survey participants in Country 2

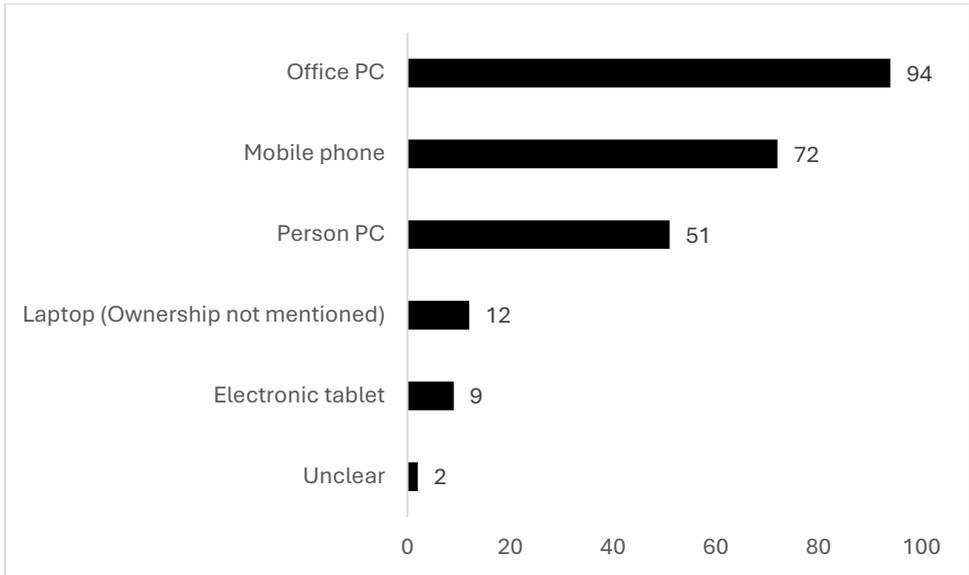

Figure 3. Electronic devices used on a daily basis by survey participants in Country 1 (multiple responses possible)

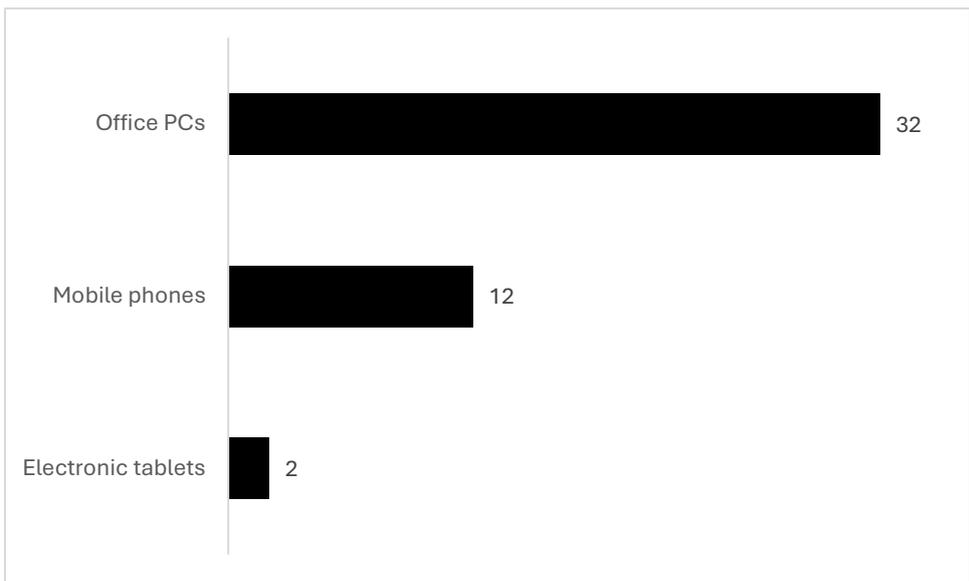

Figure 4. Electronic devices used on a daily basis by survey participants in Country 2 (multiple responses possible)

Figure 5 and Figure 6 highlight the mean scores of the two countries, while Table 4 summarises the statistically significant differences between the countries by dimension. As shown in Figures 5 and 6, the mean scores across the two countries were in general relatively high, ranging from 3.86-4.63 for Country 1 and 3.94-4.68 for Country 2. CS as a Priority, Mobile Devices, and Incident Reporting dimensions were the highest scored dimensions across the two countries. In contrast, the Integrity dimension scored the lowest in Country 1 and Password Management was lowest in Country 2.

The only significant difference between countries was the Integrity dimension (Table 4). A Mann-Whitney *U* test revealed a significant difference in the Integrity dimension between Country 1 (Mean Rank=98.38) and Country 2 (Mean Rank=135.84), *U*=1892.50, *p* < 0.001, *r*=0.36. Participants from Country 2 scored significantly higher on the Integrity dimension than those from Country 1.

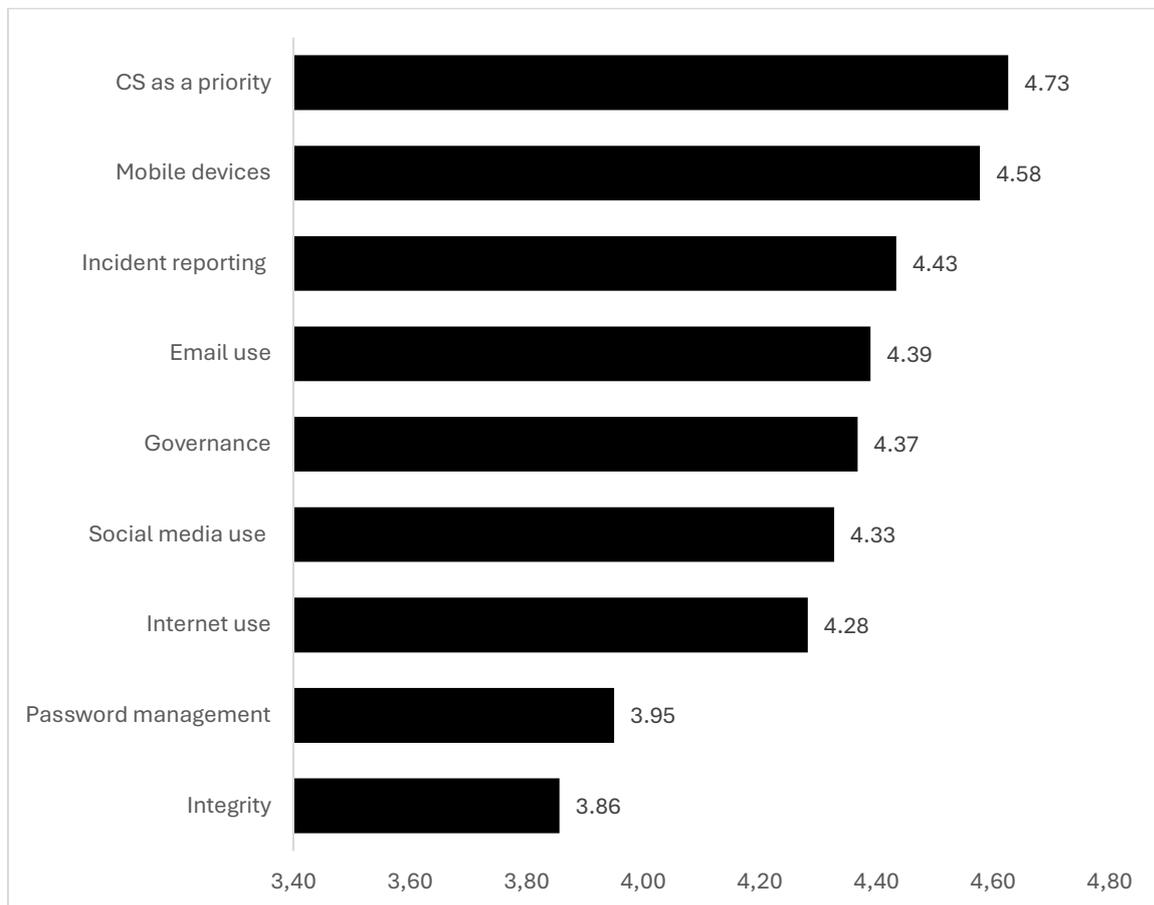

Figure 5. Country 1's survey mean scores (N=174)

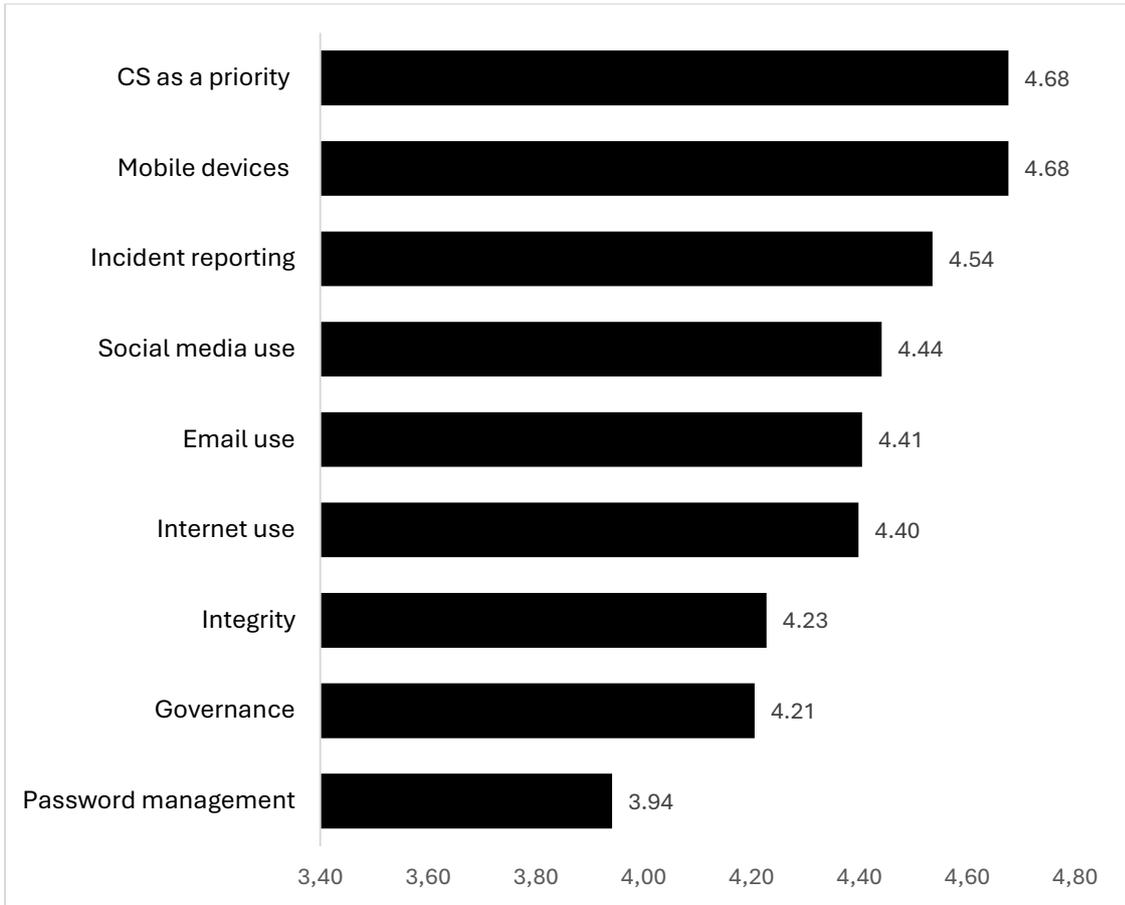

Figure 6. Country 2's survey mean scores (N=34)

Table 4. the summary of statistically significant differences by dimension

| Independent variables | Dimensions | Sub-group differences* | |
| --- | --- | --- | --- |
| | | Country 1 | Country 2 |
| Country | Password Management | No significant differences | |
| | Email Use | No significant differences | |
| | Internet Use | No significant differences | |
| | Social Media Use | No significant differences | |
| | Mobile Devices | No significant differences | |
| | Incident Reporting | No significant differences | |
| | Integrity | **Country 2 > Country 1** | |
| | Governance | No significant differences | |
| | Priority | No significant differences | |
| Employment types | Password Management | No significant differences | Unable to compute** |
| | Email Use | No significant differences | Unable to compute** |
| | Internet Use | **Permanent employees > external employees** | Unable to compute** |
| | Social Media Use | No significant differences | Unable to compute** |
| | Mobile Devices | No significant differences | Unable to compute** |
| | Incident Reporting | No significant differences | Unable to compute** |
| | Integrity | No significant differences | Unable to compute** |
| | Governance | **Permanent employees > external employees** | Unable to compute** |
| | Priority | No significant differences | Unable to compute** |
| Gender | Password Management | No significant differences | No significant differences |
| | Email Use | No significant differences | No significant differences |
| | Internet Use | No significant differences | No significant differences |
| | Social Media Use | No significant differences | No significant differences |
| | Mobile Devices | No significant differences | No significant differences |
| | Incident Reporting | No significant differences | No significant differences |
| | Integrity | No significant differences | No significant differences |
| | Governance | No significant differences | No significant differences |
| | Priority | No significant differences | No significant differences |
| Age group | Password Management | No significant differences | Unable to compute** |
| | Email Use | No significant differences | Unable to compute** |
| | Internet Use | No significant differences | Unable to compute** |
| | Social Media Use | No significant differences | Unable to compute** |
| | Mobile Devices | No significant differences | Unable to compute** |
| | Incident Reporting | No significant differences | Unable to compute** |
| | Integrity | No significant differences | Unable to compute** |
| | Governance | No significant differences | Unable to compute** |
| | Priority | No significant differences | Unable to compute** |
| Tenure at the organisation | Password Management | No significant differences | Unable to compute** |
| | Email Use | No significant differences | Unable to compute** |
| | Internet Use | No significant differences | Unable to compute** |
| | Social Media Use | No significant differences | Unable to compute** |
| | Mobile Devices | No significant differences | Unable to compute** |
| | Incident Reporting | No significant differences | Unable to compute** |
| | Integrity | No significant differences | Unable to compute** |
| | Governance | No significant differences | Unable to compute** |
| | Priority | No significant differences | Unable to compute** |
| Work context | Password Management | No significant differences | Unable to compute** |
| | Email Use | No significant differences | Unable to compute** |
| | Internet Use | No significant differences | Unable to compute** |
| | Social Media Use | **Field employees > office employees** | Unable to compute** |
| | Mobile Devices | No significant differences | Unable to compute** |
| | Incident Reporting | **Field employees > office employees** | Unable to compute** |
| | Integrity | No significant differences | Unable to compute** |
| | Governance | No significant differences | Unable to compute** |
| | Priority | No significant differences | Unable to compute** |
| Managerial roles | Password Management | **Managers > non-managers** | No significant differences |
| | Email Use | No significant differences | No significant differences |
| | Internet Use | No significant differences | No significant differences |
| | Social Media Use | No significant differences | No significant differences |
| | Mobile Devices | No significant differences | No significant differences |
| | Incident Reporting | No significant differences | No significant differences |
| | Integrity | No significant differences | No significant differences |
| | Governance | **Managers > non-managers** | No significant differences |
| | Priority | No significant differences | No significant differences |
| Nature of recruitment | Password Management | No significant differences | No significant differences |
| | Email Use | No significant differences | **Recruited through M&A processes > recruited directly** |
| | Internet Use | **Recruited directly > do not know** | No significant differences |
| | Social Media Use | No significant differences | No significant differences |
| | Mobile Devices | No significant differences | No significant differences |
| | Incident Reporting | **Recruited directly > do not know** | No significant differences |
| | Integrity | No significant differences | No significant differences |
| | Governance | **Recruited directly > do not know** | No significant differences |
| | Priority | **Recruited directly > do not know**<br>**Recruited through an M&A > do not know** | No significant differences |

*The sub-group listed before the ">" symbol scored significantly higher than the group listed after it.
** Analyses could not be conducted due to insufficient sample sizes in several sub-groups which prevented meaningful statistical computation.

## Survey results for Country 1

The only significant differences found in Country 1 were for the following independent variables: Employment types, Work context, Managerial roles, Nature of recruitment (Table 4).

**Differences by Employment Type (Permanent, temporary, or external employment)**

*Internet Use*

A Kruskal-Wallis test revealed a statistically significant difference in responses based on employment type for the Internet use dimension, $\chi^2(2) = 6.583$, $p = 0.037$. Post-hoc Dunn's tests with Holm correction showed that permanent employees scored significantly higher than external employees with a small to moderate positive effect ($z = 2.540$, $p_{(holm)} = 0.033$, $r_{rb} = 0.226$).

*Governance*

A Kruskal-Wallis test indicated a significant difference in scores across employment types for the Governance dimension, $\chi^2(2) = 7.443$, $p = 0.024$. Dunn's post-hoc tests with Holm correction showed that permanent employees scored significantly higher than external employees with a small to moderate positive effect ($z = 2.603$, $p_{(holm)} = 0.028$, $r_{rb} = 0.227$), meaning a small to moderate positive effect.

**Differences by Work Context (Field-based vs. Office-based)**

*Social Media Use*

A Mann-Whitney $U$ test revealed a significant difference in the Social Media Use dimension's mean scores between the two groups, $U = 1765.00$, $p = 0.001$, with a Hodges–Lehmann estimate of the difference in medians of -0.230 (95% CI: -0.450 to -0.110). The rank-biserial correlation was -0.341 (SE = 0.104, 95% CI: -0.508 to -0.150), indicating a moderate effect size.

*Incident Reporting*

For the Incident Reporting dimension's mean scores, the Mann-Whitney $U$ test also showed a significant difference, $U = 2061.00$, $p = 0.026$, with a Hodges–Lehmann estimate of -0.220 (95% CI: -0.340 to 0.000). The rank-biserial correlation was -0.231 (SE = 0.104, 95% CI: -0.413 to -0.031), suggesting a small to moderate effect.

**Differences by Managerial Role (Managers vs. Non-managers)**

*Password Management*

A Mann-Whitney $U$ test revealed a significant difference in the Password Management dimension's mean scores between the groups, $U = 877.50$, $p = 0.037$, with a Hodges–Lehmann estimate of -0.110 (95% CI: -0.220 to 0.00). The rank-biserial correlation was -0.306 (SE = 0.151, 95% CI: -0.547 to -0.018), indicating a moderate effect.

*Governance*

Another significant difference was observed for the Governance dimension's mean scores, $U = 714.00$, $p = 0.003$, with a Hodges–Lehmann estimate of -0.330 (95% CI: -0.670 to ≈ 0.00), and a rank-biserial correlation of -0.435 (SE = 0.151, 95% CI: -0.643 to -0.167), indicating a moderate to large effect.

**Differences by nature of recruitment (recruited directly, through an M&A process, "I do not know")**

*Perceived CS as a priority*

A Kruskal–Wallis H test showed a statistically significant difference for perceived CS as a priority across the three groups, $H(2) = 11.31$, $p = 0.004$, with a rank epsilon-squared ($\varepsilon^2$) effect size of 0.065 (95% CI: 0.044 to 0.113), indicating a small to moderate effect. Dunn's post hoc tests (with Bonferroni and Holm corrections) revealed significant differences between:

- "Recruited through M&A" and "Do not know": $z = 2.89$, $p = 0.004$, $p_{(bonf)} = 0.012$, $p_{(holm)} = 0.008$, rank-biserial correlation = 0.733, indicating large effect.
- "Recruited directly" and "Do not know": $z = 3.36$, $p < 0.001$, $p_{(bonf)} = 0.002$, $p_{(holm)} = 0.002$, rank-biserial correlation = 0.722, indicating large effect.

*Internet Use*

A Kruskal–Wallis H test revealed a statistically significant difference for the Internet Use dimension, $H(2) = 9.09$, $p = 0.011$. Dunn's post hoc tests (with Bonferroni and Holm corrections) indicated a significant difference between "Recruited directly" and "Do not know": $z = 2.58$, $p = 0.010$, $p_{(bonf)} = 0.029$, $p_{(holm)} = 0.029$, with a rank-biserial correlation of 0.677, indicating large effect.

*Incident reporting*

A Kruskal–Wallis H test indicated a statistically significant difference for the three groups, $H(2) = 11.22$, $p = 0.004$. Dunn's post hoc tests (with Bonferroni and Holm corrections) showed a significant difference between "Recruited directly" and "Do not know", $z = 2.67$, $p = 0.008$, $p_{(bonf)} = 0.023$, $p_{(holm)} = 0.023$, with a rank-biserial correlation of 0.703, indicating large effect.

*Governance*

A Kruskal–Wallis H test showed a statistically significant difference across the three employment groups, $H(2) = 6.54$, $p = 0.038$. Dunn's post hoc comparisons (with Bonferroni and Holm corrections) revealed a significant difference between "Recruited directly" and "Do not know", $z = 2.54$, $p = 0.011$, $p_{(bonf)} = 0.034$, $p_{(holm)} = 0.034$, with a rank-biserial correlation of 0.655, indicating large effect.

### 1.1.1. Survey results for Country 2

In Country 2, only the independent variable Nature of Recruitment showed statistically significant differences. A Mann-Whitney $U$ test revealed a statistically significant difference in Email Use dimension scores based on the Nature of Recruitment variable, with participants recruited directly scoring significantly higher from those recruited through M&A process ($U = 186.00, p = 0.027$). The Hodges-Lehmann estimate of the difference in medians was 0.222, with a 95% confidence interval ranging from $1.24 \times 10^{-5}$ to 0.445. The rank-biserial correlation was 0.470 (SE = 0.211), indicating a moderate effect size, with a 95% confidence interval from 0.089 to 0.731. This means that those recruited via M&A processes scored significantly higher than those recruited directly into the organisation for the Email Use dimension. There were no other statistically significant differences ($p > 0.05$) for other dimensions because effect sizes were small, and confidence intervals included zero.

No significant differences were observed regarding Gender or Line Managerial Responsibility. It was not possible to conduct analyses for the remaining independent variables, including Employment Type, Age Group, Tenure, and Work Context, due to insufficient sample sizes in several sub-groups, which prevented meaningful statistical computation.

### Stage 2: Interview findings

For Country 1, we interviewed a total of 20 employees: 13 through video calls on Teams and 7 through written chat on Teams. For Country 2, a total of 10 interviews were conducted, in which nine were video-interviewed and one on the Teams' chat. All interview participants were internal employees. We did not interview any external employees because they did not respond to interview invitations and informal chats. Table 5 summarises the key interview findings by country. In general, there were similarities in the interview responses across themes.

Table 5. Summary of key interview findings by country

| Interview themes | Country 1 | Country 2 |
|---|---|---|
| Understanding of CS | CS was generally understood as protection and security of both personal and organisational information, data, systems, networks and devices. | CS was generally understood as safeguarding our systems, data, and customer information from unauthorised access, cyber threats, and malicious attacks. |
| CS as a priority | CS was generally perceived as a priority in their team, department, management, or business area, demonstrated through various CS measures and initiatives. | CS was generally perceived as a priority in their teams, departments, management, or business areas, demonstrated through various CS measures and initiatives. |
| Current CS state in the organisation | The overall CS state was generally perceived as positive, proactive, and vigilant, demonstrated by examples of initiatives and activities by interview | The overall CS state was generally perceived as positive, proactive, and vigilant, demonstrated by examples of initiatives and activities by interview |
| | participants (e.g. CS moments, phishing campaigns and simulations, technical measures). Some improvement areas were also mentioned. | participants (e.g. mandatory regular e-learning, email classification) |
| CS challenges | Challenges mentioned related to traveling, technical, e-learning issues, varied CS awareness, external employees, and external factors such as workload and deadlines. | There were perceived discrepancies between what was understood as CS policies (i.e. no links in email policy) and what was being practiced (i.e. still receiving legitimate links in emails from the organisation). |
| CS related policies and procedures | While generally perceived as useful, interview participants varied in their knowledge of and familiarity with the content and location of these documents. | While generally perceived as useful, participants varied in their knowledge of and familiarity with the content and location of these documents. |
| Mandatory CS e-learning | E-learning was generally perceived as sufficient, understandable and relevant. However, some perceived that the CS evolved, and thus training should be frequently updated, and too much information might lead to information overload. | The e-learning was generally perceived as useful, sufficient, and raising CS awareness. |
| Perceptions about phishing | Interview participants recognised the importance of identifying phishing attempts and described how these could be detected and handled. | Interview participants recognised the importance of identifying phishing attempts and described how these could be detected and handled. |
| Perceptions about the CSC survey | • Those who recalled the survey gave no comments or positive feedback.<br>• Some thought that the survey was phishing simulation and reported it. | • Those who recalled reported the survey email invitation as a phishing attempt due to the country's no links in email policy.<br>• Those who participated mentioned the survey was too long and repetitive.<br>• All were interested in the findings. |
| CS Reporting | All interview participants indicated that they would feel comfortable reporting CS incidents or concerns via several channels: namely, approaching their line | All interview participants indicated that they would feel comfortable reporting CS incidents or concerns via several channels, namely, approaching their line |

|  | managers, contacting the IT department, or using the "Report Phishing" function in Microsoft Outlook. In practice, however, most had only ever reported phishing attempts. They were also unsure how to go about reporting and need guidance on what is worth reporting and how to report non-phishing incidents. | managers, contacting the IT department, or using the "Report Phishing" function in Microsoft Outlook. In practice, however, all had only ever reported phishing attempts. They were also unsure how to go about reporting and need guidance on what is worth reporting and how to report non-phishing incidents. |
|---|---|---|
| The role of access control | Document classification was perceived as important and functional, but also made it difficult to get access to systems from other devices. Interview participants had varied experience and knowledge with using it. | Document classification was perceived as useful, straightforward, and important. However, participants sometimes forgot to apply it, were unclear about the difference between "confidential" and "secret," and faced challenges sharing "confidential" documents with external parties. |

## Interview findings for Country 1

*Understanding of CS*

Interview findings revealed that participants broadly understood CS as encompassing the protection of information, data, systems, networks, and/or devices. Interview participants were generally aware that the use of multiple digital tools would increase their exposure to cyber threats, making proper handling of data and information a critical factor in reducing risk. Some also described CS as safeguarding both personal and organisational data, with specific threats to focus on including phishing emails and malicious QR codes. Some mentioned CS consequences of clicking phishing links or providing credentials, such as financial fraud, data breaches, and business disruptions. Moreover, one interview participant highlighted the connection between CS and the physical and operational safety of individuals and organisations. Nevertheless, some interview participants primarily associated CS with the ability to identify phishing attempts.

*CS as a priority*

Interview participants generally perceived that CS was prioritised within their teams, by management, or within their business areas. This prioritisation was evident in mandatory CS e-learning, optional training programmes, encouragement to report CS incidents, regular phishing simulations, internal sharing of CS-related information, implementation of access control, multi-factor authentication, secure login mechanisms, and the designation of super users.

Some interview participants mentioned that the mandatory CS e-learning was likely to reinforce favourable CS practices. Additionally, some saw sharing CS information within teams and with managers – such as through CS moments, short and practical CS tips typically shared in meetings, presentations, and flyers – as raising awareness and directly influencing day-to-day behaviours to increase caution when handling email links.

Specific for external employees, CS was said to be prioritised through reinforcement to use official organisational email addresses rather than personal ones. This was because these employees might not necessarily receive the official organisation's PCs or might use multiple PCs from different organisations and/or personal PCs.

*Current CS state in the organisation*

Interview participants identified current CS measures in place at the organisation. Examples included secure login procedures, automatic activation of VPN once outside internal network, multi-factor authentication, phishing simulations, mandatory CS e-learning, CS awareness campaigns, and CS moments in meetings. Participants said these efforts contributed to a perception that the organisation was proactive and vigilant in its CS work.

However, participants also mentioned several areas for improvement. These included a lack of prompts to regularly change passwords, too limited basic CS training for external employees, and the use of personal (unmonitored) devices by external employees who often lacked access to internal systems. In addition, interview participants highlighted that as new technologies were introduced, the existing lines of defence must evolve to address emerging risks. Variations in CS exposure based on role, such as auditors navigating different network environments, further emphasised the need for tailored and consistent CS strategies across the organisation.

*CS challenges*

Interview participants mentioned several challenges to prioritizing CS: traveling, use of mobile phones, external factors, external employees' limited access to the organisation's internal platforms, and limited CS training. Some interview participants mentioned that roles involving frequent travel, such as auditors, faced heightened risks due to the increased need to work in off-site or public places. The mentioned risks included shoulder surfing and the increased use of public or customer networks in airports, hotels, or client offices. Participants also identified the use of mobile phones as a potential vulnerability, especially when communication took place through unmonitored channels such as SMS or personal emails. They also noted that current mandatory CS e-learning often failed to address using multiple device types in a secure way.

According to participants, external factors such as time pressure, conflicting priorities, and customer demands sometimes led to more difficulties in prioritising cybersecure practices. Moreover, differences in provided CS training and access or lack thereof to the organisation's digital internal platforms between internal and external employees created gaps in CS awareness and compliance between them. Specifically, participants mentioned that external employees often lacked access to the organisation's official PCs, making it difficult to monitor activity or ensure consistent favourable CS practices across the workforce. They also highlighted the need for more tailored CS training and improved support for roles with varying CS risk profiles such as field- vs. office-based employees.

*CS related policies and procedures*

Interview participants had varied levels of familiarity with CS-related policies and procedures within the organisation. While some were unfamiliar altogether, others knew only the procedures relevant to their daily work or those covered in training or felt they would know where to find them should it be needed. Only a few said they had read the relevant ones. The perceived accessibility also varied, in which some found the documents easy to locate, while others found them scattered and hard to find or said there were other conflicting priorities to regularly looking them up.

Those who had accessed the CS related policies and procedures documents perceived them as positive and appreciated that they were usually short (1–2 pages), clearly written, and organised with bullet points or simple steps. Specific examples such as procedures for document classification for access control and identifying phishing emails were seen as especially helpful and actionable. However, several noted the possibility of information overload and suggested that periodic sharing of key extracts would be more effective than having to search through extensive documentation.

*Mandatory CS e-learning*

Interview participants generally found the mandatory CS e-learning to be sufficient, understandable, and relevant. They appreciated the use of examples and phishing simulations which helped make the material more relatable and practical.

However, some noted that CS was a constantly evolving area; training needed to be updated frequently to stay relevant and not only focused on phishing identification and handling. There were also concerns about information overload that highlight the need for more focused and digestible content. Suggestions for improvement included offering regular refreshers, increasing the frequency of critical modules, and making CS e-learning more engaging by incorporating workshops, real-life scenarios, and gamification. Interview participants also recommended tailoring content to different regions and roles, sharing best practices, providing details on emerging risks, and using more interactive formats.

*Perceptions about phishing*

Interview participants described a range of strategies they used or would use to identify phishing emails, such as checking the sender's email address, evaluating the content, hovering over links, and paying attention to caution messages, typos, or unusual branding and logos. They were also cautious of emails requesting data or creating a sense of urgency. A key takeaway mentioned by interview participants was the need to be extra vigilant when feeling tempted or rushed, as that's when people were most vulnerable. Still, several noted that real-world, sophisticated phishing attempts might be difficult to spot.

When suspecting a phishing attempt, interviewees mentioned common actions like permanently deleting the email, using the report phishing button on Microsoft Outlook, or notifying a line manager. If the suspicious message appeared to come from a customer, some would follow up by phone to confirm the sender's identity.

*Perceptions about the CSC survey*

Interview participants who participated in the CSC survey generally gave positive feedback on the survey that was sent prior to the interviews and did not raise major concerns. However, they shared a few improvement suggestions in the interview. Some participants initially mistook the survey for a phishing attempt due to the email containing a link, inconsistent fonts and colours, and the absence of the organisation's logo.

Additional feedback included suggestions to focus more on realistic scenarios that could impact the organisation directly and to expand the survey to cover aspects such as CS training and how to securely use multiple devices. These comments highlighted opportunities to strengthen the clarity and relevance of future surveys as well as how to best disseminate the survey.

*CS Reporting*

Most interview participants said they would feel comfortable and safe reporting CS incidents, particularly phishing, and emphasised the importance of doing so. They highlighted the "report phishing" button in Microsoft Outlook as being especially useful, and the majority were familiar with this reporting method. Some also mentioned raising issues with the IT department or consulting line managers as other available channels to report CS incidents.

However, the majority of the interview participants noted that their knowledge was largely limited to reporting phishing emails and they were unsure how to report other types of incidents or what is worth reporting as a CS concern. The majority of interview participants had never reported CS incidents other than using the report function on Microsoft Outlook.

Some mentioned the need for more consistent, clearer guidance and a more coordinated reporting process. Suggestions for improvement included simplifying the reporting of non-phishing concerns, offering clearer guidelines on what to report where, and providing more encouragement to report anything in general. Some also pointed out that the abundance of information can be overwhelming and make it difficult to effectively navigate reporting procedures.

*The role of access control*

Interview participants generally viewed access control as important and effective for protecting systems and data. This was done by classifying documents into different levels of accessibility, for example, "available to all" through to "available only to those allowed". Multiple-factor authentication was seen as a beneficial layer of security. However, some noted that access controls could create practical challenges, such as difficulty accessing documents stored in the systems from different devices and by external collaborators from other organisations.

When it came to document classification, experience and awareness varied. Some participants were unfamiliar or had only heard of classification possibilities, while others had a clearer understanding of the classification levels. Those with experience found it useful for safeguarding sensitive information. A suggested improvement was to implement automated support – such as having the computer suggested appropriate document labels based on the content's sensitivity – to make classification easier and more consistent.

### Interview findings for Country 2

*Understanding of CS*

Interview participants generally understood CS was about safeguarding the systems, data, and customer information from unauthorised access, cyber threats, and malicious attacks. They emphasised safeguarding confidential information so that only authorised individuals, both inside and outside the organisation, could access it.

Additionally, interview participants recognised CS as a defence against various threats such as phishing, hacking, and extortion. They also highlighted that CS meant constant vigilance, being able to spot suspicious emails, avoid harmful links, and maintain strong awareness to minimize risks and help keep the organisation secure.

*CS as a priority*

Interview participants generally agreed that CS was a priority within their teams and departments or business areas. They highlighted various practices that supported this, such as enforcing use of access control, following secure coding standards, holding CS moments in meetings, sharing concerns among colleagues, and participating in CS e-learning and training. Reminders to stay vigilant, such as being cautious against clicking suspicious links and raising issues to the IT department, were also common ways their teams prioritised CS.

However, several interview participants suggested the need to be more consistent in daily CS practices across the organisation and more effective ways to raise CS awareness. Busyness and competing workloads were cited as barriers to consistently prioritising CS. Additionally, interview participants suggested the need for better monitoring, more integrated security tools in daily workflows, and improved access to these resources to strengthen the organisation's overall CS state.

*Current CS state in the organisation*

The organisation's current CS approach was generally viewed as proactive and effective. Mandatory regular e-learning and phishing simulations were highlighted as key practices that help employees stay vigilant. Other positive measures included secure storage, secure sites, password protection, and document classification which participants said protect both digital and physical information. Some also mentioned that within their teams they held CS moments to share lessons learned from incidents or concerns to reinforce awareness and encourage employees to prioritize CS.

However, participants noted some inconsistencies in practices across regions and departments. For example, while Country 2 decided to have a "no-links in emails" policy, employees still received legitimate emails containing links from the organisation from other regions and/or departments, even from senior managers, causing confusion and mixed reactions. There were also concerns about using email to share information instead of more secure internal platforms which were described as cumbersome due to access control restrictions. Participants suggested improving consistency between the organisation's policies and actual practices as ways to strengthen the organisation's CS state and balance usability and security.

*CS challenges*

While a few interview participants reported no difficulties in prioritising CS within their teams, some reported that a lack of CS awareness and vigilance among employees was a significant challenge. Participants said this was partly due to insufficient and irrelevant CS training and partly the result of complacency among some employees who did not perceive cyberattacks as a real threat to the organisation.

In general, interview participants reported CS challenges as inconsistencies in how CS was practiced and communicated, with discrepancies between policies and actual behaviours. Competing priorities such as time pressures, understaffing, customer demands, and costs also made it difficult for some teams to consistently prioritize CS. Participants described guidance and information on CS as fragmented and hard to find, adding to the challenge. They saw sharing information using internal digital platforms as cumbersome, specifically with external collaborators from other organisations. Specifically, giving access to external organisations, such as customers, required approvals and multiple logins and slowed down collaboration processes. Many emphasised the need to strike a better balance between CS and productivity or efficiency.

*CS related policies and procedures*

Interview participants expressed varied familiarity with CS-related policies and procedures. Many were unsure exactly where to find these documents but were relatively confident about how to look for them. Some noted that the intranet's search functionality declined, making it more difficult and time-consuming to locate specific information unless they knew exactly where to look. Others admitted they rarely or never actively sought out these documents.

While some found the policies and procedures easy to access and useful – particularly for handling incidents, sharing information with customers, and gathering relevant details – they often forgot about them or did not use them regularly. To improve accessibility, participants suggested making CS policies more visible and consolidated, such as by placing them in prominent, relevant locations like a dedicated cyber page or security hub. They also recommended periodic reviews to ensure the policies remain up to date and relevant. Overall, the policies were seen as valuable resources when needed but could benefit from better visibility and easier access.

*Mandatory CS e-learning*

Interview participants generally found the mandatory CS e-learning to be sufficient and useful, especially when the modules were short and included scenario-based examples focused on phishing identification and basic security practices. Many felt the e-learning helped raise awareness, although one interviewee felt their training was insufficient. Several interview participants pointed to another mandatory CS training, designed specifically for software developers, as a good example of how the general mandatory e-learning could be improved, noting that the developer training included multiple levels of difficulty.

Suggestions for improving the mandatory CS e-learning included making it more interactive and engaging with less text, incorporating more tests and dilemma-based scenarios, and providing feedback on performance. Interview participants also recommended offering incentives, using realistic and relatable examples, and developing role-specific training tailored to different job functions. In addition, there was a call for more training focused on emerging CS threats and practical ways to handle them.

*Perceptions about phishing*

Interview participants described several strategies they use to identify phishing attempts, including hovering over links and sender names, verifying the sender's company, checking for spelling or grammar errors, and being cautious of emails that create a sense of urgency or ask for sensitive details. Many also stressed the importance of confirming if an email was expected or relevant and whether they recognised the sender. When unsure, some reported contacting the IT department for advice.

Regarding phishing simulations, some interview participants felt the tests could be made more sophisticated to better reflect the complexity of real-world attacks. Several preferred accessing systems directly via the intranet rather than clicking links in emails. A few shared experiences of phishing attempts or impersonation scams targeting their email addresses or customers. When confident an email was phishing, interview participants typically reported it using the "report phishing" button in Microsoft Outlook. Such reporting phishing function was mentioned by all participants as very easy and user-friendly as it was only "one-click away".

*Perceptions about the CSC survey*

Some interview participants did not take or did not recall participating in the CSC survey. A few who did, reported the survey email as spam, partly due to the country's policies discouraging clicking on email links. For those who participated, the survey was described as too long and repetitive, with suggestions to allow saving progress and to better target questions to participants. Despite these issues, there was notable interest in receiving and learning from the survey findings.

*CS reporting*

All interview participants expressed feeling comfortable reporting CS-related incidents or concerns. Common reporting methods included contacting their line managers, reaching out to the IT department or IT security teams, using the "report spam" button in Microsoft Outlook. Nevertheless, only a few had reported non-phishing CS concerns or incidents, mainly because they did not feel they had encountered non-phishing CS incidents.

*The role of access control*

Interview participants generally found the mandatory e-learning on access control and document classification to be useful, practical, and easy to understand. Document classification was viewed as straightforward and important, particularly for protecting confidential information before uploading or sharing it on platforms. However, they mentioned a key challenge was that labelling emails as "confidential" sometimes prevented external recipients, such as customers, from accessing them. Another common issue was forgetting to classify documents, especially when the task was infrequent. In addition, some interview participants were unclear about the distinctions between different classification levels "confidential" and "secret."

## Improvement suggestions

We analysed participants' suggestions for improving the organisation's CS gathered from open-ended survey responses and interview discussions together to identify common themes and insights by country. Table 6 shows the high-level summary of key improvement suggestions by survey and interview participants. See Appendix 4 for the more detailed suggested improvements. In general, the participant suggestions across the two countries were very similar if not identical in some themes. This consistency in responses was particularly striking given that the qualitative analyses of suggestions from survey and interview participants were conducted separately for each country. This alignment suggests that, despite contextual differences, participants across both countries face similar CS challenges and recognize common areas for improvement. This points to potentially shared organisational or sectoral dynamics.

The only differences observed in the improvement suggestions between the two countries were that participants from Country 2 were more focused on reinforcing consistent CS policies and practices across regions, with particular emphasis on the use of links in emails. In contrast, participants from Country 1 focused more on improving CS for external employees who might not have received the organisation's official PCs, access to the intranet and CS information, or the same CS e-learning as internal employees. These differences likely reflect contextual and cultural factors specific to each country. While CSC tends to involve many shared organisational challenges, such as the need for clear policies, awareness, and training, the way participants experienced and prioritized these challenges could differ due to local organisational structures, employment practices, and cultural attitudes toward communication, hierarchy, and responsibility. Therefore, although there was substantial overlap in the suggested improvements, local context remained an important factor in shaping the focus of CS initiatives.

Table 6. High-level summary of key improvement suggestions by country

| Improvement suggestion themes | Country 1 | Country 2 |
|---|---|---|
| Leadership and CSC | Leadership commitment and leaders promoting strong CS culture | Improve leaders' CS competence and make leaders to be on the same page to promote strong CS culture |
| Organisational learning | Foster organisational learning and awareness initiatives for all internal and external employees | Foster organisational learning and awareness initiatives |
| CS governance | Establish CS governance and measures and ensure consistent CS practices across the organisation, accessible to all internal and external employees | Establish clear and consistent CS governance, particularly in use of links in emails |
| CS training | Make trainings effective, continuous, interactive, targeted, tiered, and available for all (internal and external) employees | Make trainings tiered, targeted, realistic, and interactive |
| CS reporting | Make reporting for non-phishing easy, encouraged, and guided | Encourage and ensure employees know what how and where to report CS concerns, near-misses, and incidents |
| CS technical measures | Strengthening CS technical measures | Improve CS technical infrastructure, password management, and access control |

# Discussion

We conducted a CSC mixed-methods assessment combining survey and interview data across selected units in two countries. This work is part of a broader organisational initiative to improve internal CS. The survey results indicate generally high levels of CS awareness among survey participants from both countries, with mean scores across dimensions ranging from 3.86 to 4.68 on a 5-point Likert scale.

## Minimal country-level differences

There were relatively high mean scores across dimensions without many statistically significant subgroup differences across countries or within country demographic categories. The only statistically significant difference observed between two countries was that Country 2's survey participants demonstrated better understanding that no digital platform can be considered 100% secure (Integrity dimension). One possible explanation for this result is that all survey participants from Country 2 were permanent employees. Their employment status likely provides them with greater and more consistent exposure to the organisation's internal CS initiatives, policies, and training, which may have contributed to their stronger awareness in this area. The lower response rate in Country 2 (30%) may also have skewed results if survey participation was higher among CS champions. This suggests that CSC is more similar than different across the two countries or that high variability in the data masks underlying differences. Such quantitative results benefit from qualitative results to understand the survey results better, as also suggested in previous studies [27, 28].

Interviews showed both similarities and differences across the two countries. Specifically, Country 1 expressed greater concerns about external employees' CS risks and Country 2 maintains a policy of not using links in emails. Country 1's concerns seem to stem not from external workers themselves but from their likely access to sensitive customer information, their limited or different CS training, limited access to the organisation's official PCs, CS policy and guidance.

Importantly, some interview participants from Country 2 revealed that they reported the CSC survey email invitation as a phishing attempt because it contained a link in violation of their no-link in emails policy. This likely explains the low response rate to the survey in Country 2. This suggests a limited association between phishing simulation test results and broader CSC. In other words, stronger performance on phishing simulation tests may not necessarily indicate a more robust CSC. Instead, it could simply be the result of specific policies, such as a strict "no links in emails" rule. With such a policy in place, employees may begin to disregard the legitimacy of links by default and choose to either ignore or report all emails containing links as phishing, rather than assessing them. Our findings highlight the importance of not relying solely on phishing simulation results to understand CSC, as such tests are likely to capture only a narrow aspect of a much more complex, multifaceted picture [29]. In addition, the perceived user-friendliness of the reporting phishing button on Microsoft Outlook is likely to contribute to the compliance of the no-links in email policy. This finding is supported by a recent study that has found that persuasive technological solutions that are straightforward and user-friendly motivate employees to report phishing emails [29].

In contrast, the higher survey response rate in Country 1 (67%) may reflect a stronger willingness to share their perspectives about CS. The large proportion of external employee participation (55%), indicating a broader and more diverse group of employees, could also help explain why phishing simulation test results in Country 1 were generally lower compared to other countries. These findings point to an opportunity to foster more inclusive CS dialogues that actively engage different internal and external employee groups.

### Strong areas identified in the interviews in Country 1 and 2

Similar to survey findings, interview findings show minimal differences between the two countries (Table 5). Specifically, interviews show several strong areas in the organisation's CSC across the two countries. Interview participants described CS as a clear organisational priority, grounded in concrete, observable practices like regular phishing simulations and campaigns, implementation of technical safeguards such as automatic VPN, access control measures, mandatory training, and CS moments. Similarly, interview participants described the organisation's CSC as proactive, vigilant, and maturing and mentioned the organisation's combination of technical measures, training, and campaigns as activities that raise their awareness.

In addition, participants said the mandatory CS e-learning is useful, sufficient, and understandable. They showed a high awareness of phishing threats, which interviewees attributed to the training and practical exposure through phishing campaigns and simulations. Interview participants explicitly expressed comfort in reporting CS concerns and incidents.

These observed similarities between countries, despite differences in survey response rates and demographic compositions, highlight a positive outcome of global CS measures such as the phishing identification training and campaign. It is important to note that we only interviewed internal employees, who may be more familiar with and supportive of these initiatives, which could have influenced the results in a more positive direction. Nevertheless, in general, interview findings show that while local engagement patterns may vary, the core global CS messages and initiatives, especially those related to phishing identification, are resonating across the two countries.

### Key findings specific to Country 1

In addition to the overarching similarities across countries, the survey and interview revealed several findings specifically within Country 1. In brief, survey participants who identified themselves as permanent employees, hold a managerial role, or field-based employees reported higher scores in several different dimensions of the CSC survey (Table 4). These groups may already demonstrate stronger CSC practices, making them potential change agents as well as have different needs for training. Tailoring CSC improvement efforts thus needs to consider employment and managerial roles of the target employees. We discuss the key findings below.

*First*, permanent employees demonstrated higher awareness of the Internet Use and Governance dimensions compared to external employees. This means that permanent participants are likely to exhibit a stronger understanding of risks associated with downloading files and entering information online, navigating external websites, and greater familiarity with the organisation's CS policies and procedures. These findings may be explained by the fact that permanent employees have access to the organisation's CS policies and procedures and receive more mandatory CS e-learnings than the external employees. In addition, permanent employees are likely to receive more exposure to internal communication about CS, such as through CS moments, raising their awareness of the risks associated with using the internet in general.

*Second*, field-based employees scored higher in both Social Media Use and Incident Reporting dimensions. Field-based employees are typically auditors, assessors, or surveyors whose work focuses on identifying compliance or non-compliance with specific standards and requirements. These findings may indicate that hands-on experience in variable field environments contributes to a transferable awareness of risk in CS, particularly regarding interactions on social media platforms and the importance and procedures for reporting incidents.

*Third*, managers outperformed non-managers in both the Password Management and Governance dimensions. These findings likely reflect greater responsibility and access to information, as well as stronger alignment with organisational CS expectations. For example, managers demonstrate higher awareness of what constitutes strong vs. weak passwords and greater perceived familiarity with internal CS procedures. This suggests that managerial knowledge and responsibilities are likely transferable to CS.

*Fourth*, we found no statistically significant differences across age groups, gender, or tenure in the organisation. Future studies with larger populations are required to assess whether the finding results from variability in responses or from these factors having limited influence on CSC.

## Improvement areas identified in Country 1 and 2

While the overall CSC appears positive and proactive, our findings also reveal several challenges and improvement areas, particularly related to role clarity, information accessibility, and scope of awareness across two countries.

*First*, line managers did not perform significantly better or worse than non-managers in most dimensions in the survey. While managers in Country 1 showed significantly higher survey scores on the Governance and Password Management dimensions, this pattern did not hold for the remaining seven dimensions, or, for managers in Country 2. This is particularly concerning since line managers are consistently treated as the first point of contact for all CS concerns, near-misses, and incidents for internal as well as external employees. In addition, they are responsible for overseeing their team members' CS behaviours. Importantly, interview findings revealed uncertainty and unclarity around what happens or what would happen to the reported CS concerns or incidents, suggesting a gap in feedback loops. Previous studies have pinpointed the importance of top management's support, involvement and commitment if CSC is to be improved [28]. Such commitment needs to be trickled down by supporting line managers with sufficient CS competence within their roles and fields.

*Second*, an overemphasis on phishing was a recurring theme in interviews. While important, this focus may unintentionally limit participants' understanding of CS risks. The success of the organisation's phishing identification initiatives and campaigns likely contribute to the beliefs that all links are malicious or that CS concerns begin and end with phishing detection. In addition, when participants encounter non-phishing-related concerns, they contacted their line managers but remained unsure about what actions this led to. This highlights a critical need for clearer guidance, defined escalation procedures, and role-specific training particularly for line managers [30].

*Third*, interview participants, particularly for those in Country 2, were confused about the use of links in emails in the organisation. Specifically, Country 2's interview findings revealed a lack of uniformity in the organisation. This highlights the importance of harmonising CS policies and communication practices across the organisation's regions and business areas. The importance of having clear policies and procedures that are well communicated and understood by employees has been suggested the top 2 factor to build and maintain a CSC [28].

*Fourth*, participants had considerable variation in perceived familiarity with the organisation's CS related policies, procedures, and guidelines. This finding is similar to previous studies [28]. Interview participants expressed different levels of awareness about where to find these resources and what constitutes a reportable event or concern. Most participants reported having submitted only suspected phishing emails, with limited understanding of how to escalate or report other types of CS concerns or incidents. This indicates a need to clarify the scope of CS threats beyond phishing and to improve navigability and consolidation of CS related information across platforms.

*Fifth*, the interviews revealed several organisational constraints that limit CS engagement. These include competing priorities, workload, time pressures, frequent travel, understaffing, and cost concerns. These factors contribute to a tendency to deprioritize CS [28, 31].

*Sixth*, we identified a specific challenge in Country 1 related to external employees: external employees have limited access to internal secure platforms, the organisation's official computers, CS policies, procedures, guidelines, and e-learning resources. Due to their external status and use of personal devices, it is very difficult or nearly impossible to monitor CS activities and ensure compliance with the organisational standards. Interview participants emphasize these concerns, highlighting that external employees frequently handle sensitive information and interact directly with customers, which increases the risks [28]. These findings highlight the need for tailored strategies to extend CS protections and training to external employees, including solutions for secure device use, improved access, and enhanced monitoring where feasible [30].

## Key recommendations to improve CSC in Country 1 and 2

Based on the survey and interview findings (Table 4 and 5), as well as the improvement suggestions by the survey and interview participants (Table 6 and Appendix 4), we propose and prioritize the following key recommendations to improve the CSC in the two countries.

**First, foster a culture where everybody shares ownership and responsibility in CS.** This can be achieved by focusing on practices to strengthen leadership commitment and employee engagement.

- *Leadership commitment*: (1) enhancing senior and line managers' CS literacy and aligning their CS priorities, (2) appointing CS leads as dedicated points of contact by region and business area, (3) implementing routine discussions led by line managers to emphasize individual ownership in CS, (4) encouraging leaders to serve as role models in promoting CS best practices.
- *Employee engagement*: (1) providing regular updates on real-world and the organisation's cyber threats, emphasising on what and how it happens and potential consequences, (2) conducting frequent CSC assessment, (3) implementing mandatory CS moments in townhall and departmental meetings, (4) providing feedback on phishing test performance and reported emails, and (5) establishing reward and recognition initiatives for proactive CS efforts.

***Second***, **ensure consistency in CS policies and practices across regions.** This can be achieved by: (1) aligning policies with real-world practices through periodic reviews and clear, accessible guidelines, (2) periodically reviewing the relevance of existing policies and procedures, (3) making CS information accessible to both internal and external employees, (4) ensuring consistency in use of links across regions, (5) consolidating all CS information and guidance in one place, and (6) establishing and communicating clear CS emergency plans and playbooks.

***Third***, **tailor training to ensure all employees understand what CS issues are worth reporting and how to report them effectively.** This can be achieved by: (1) encouraging reporting of all concerns and near-misses, not just incidents, (2) making it easy to report more than just phishing.

***Fourth,*** **design CS training to be inclusive, engaging, and relevant.** This means ensuring it reaches both internal and external employees, is tiered and role-based (e.g. field vs. office-based, managers vs. non-managers), uses real-world cases, blends e-learning with classroom sessions, and is delivered interactively and continuously.

## Limitations

Our study has several limitations. *First*, the response rate varied significantly between Country 1 and 2, with a significantly lower rate in Country 2. One contributing factor to this may have been that the survey was distributed via a link in an email, and it was discovered during the study that Country 2 had a "no links in emails" policy. This varying response rate may introduce non-response bias, particularly if those who chose to participate differ systematically from those who did not (e.g. CS champions being more likely to respond). *Second*, external employees were not represented in Country 2, limiting our ability to compare perceptions and behaviours between employee types across both settings. *Third*, although the survey instrument was designed to cover key dimensions of CSC, it captures self-reported perceptions rather than actual behaviours, which may be influenced by social desirability bias. While we complemented survey results with qualitative interview data, the number of interviews was relatively small and may not fully capture the diversity of experiences within each country or business area. *Fourth*, Country 2's analyses for Employment Type, Age Group, Tenure, and Work Context were not conducted due to small subgroup sizes, limiting our ability to detect potential differences related to these variables. *Fifth,* our study focused on a limited number of units and may not generalize to the entire organisation. Future studies could benefit from broader sampling, longitudinal tracking, and integration of behavioural data (e.g. phishing simulation test results) to build a more comprehensive picture of CSC and the effectiveness of CSC improvement efforts over time.

# Conclusions

Our mixed-methods assessment provides a more detailed understanding of CSC across selected units in the two countries. While survey results show generally high levels of CS awareness and commitment, they also reveal important local variations influenced by employee roles, country-level organisational policy, and engagement patterns. Importantly, interview findings offer valuable context to further interpret survey findings, such as concerns about external employees' access and the varying impact of a "no-link" policy. These insights emphasize that even where overall awareness seems to be high at first glance, specific local factors continue to shape how people understand an enact CS. The findings thus highlight that one size does not fit all and the importance of tailoring improvement efforts, training, and communication strategies to different employee groups and the organisational culture in different regions.

Future work should prioritize scaling the CSC assessment to cover the entire organisation to enable a more consistent and organisation-wide strengthening of CSC. A broader assessment will offer a more representative and comprehensive picture of the organisation's CSC, allowing for more targeted and effective improvement efforts across all units and geographies. Integrating objective measures, such as phishing simulation tests, into the CSC assessment is also crucial. These measures can provide behavioural evidence that validates or contrasts self-reported assessment, enabling a deeper understanding of human factors in CS. In addition, future efforts should evaluate the impact of tailored improvement efforts on CSC outcomes. Systematically tracking which strategies lead to measurable improvements allows the organisation to identify the most effective ones, allocate resources more efficiently, and continuously refine its approach.

Addressing local challenges while maintaining strong leadership and high engagement is key to strengthening the organisation's overall CSC. This approach is particularly vital in safety-critical industries and organisations supporting critical infrastructure where the human factor plays a central role in preventing incidents. By strengthening its CSC, the organisation can transform employees from being a point of vulnerability into an active line of cyber defence, complementing technical safeguards and building cyber resilience across diverse operational settings.

# Acknowledgement
Special thanks to Sofie Alvestad Forfang for her support with the interviews, to the Country Chairs and managers across the participating units for enabling the study, and to all participants for their valuable contributions.

# Funding
This study was internally funded by DNV as part of its Cybersecurity Game Changer programme. No external funding was received.

# Conflict of interest
The authors declare no conflicts of interest.

**List of Appendices**

# Appendices

Appendix 1. Demographic questions for the survey

1. I am based in
    a. Country 1
    b. Country 2
2. My employment is:
    a. Permanent employee
    b. Temporary employee
    c. External employee
3. Gender:
    a. Female
    b. Male
    c. Prefer not to say
4. My age group is:
    a. >25 years old
    b. 25-30 years old
    c. 31-40 years old
    d. 41-50 years old
    e. >50 years old
    f. Prefer not to say
5. I have been working in the organisation for:
    a. <1 year
    b. 1-5 years
    c. 6-10 years
    d. 11-20 years
    e. 21-30 years
    f. >30 years
6. I am predominantly:
    a. A field employee (e.g. surveyors, assessors, auditors)
    b. An office-based employee
7. I am a line manager (people reporting to me):
    a. Yes
    b. No
8. My primary role is: free text
9. I am recruited directly by the organisation (not through a M&A process):
    a. Yes
    b. No
    c. I don't know
10. On a daily basis, I usually work using (multiple responses possible):
    a. The organisation's PC
    b. Mobile phone

  c. Electronic Tablet
  d. .. free text
11. My business area has prioritised raising the cybersecurity awareness in the past year.
  a. Strongly agree
  b. Agree
  c. Neutral
  d. Disagree
  e. Strongly disagree
12. In my opinion, actions or initiatives that can improve our organisation's cybersecurity culture are: …. Free texts…..

Appendix 2. All questionnaire items and their associated dimensions

| Questionnaire | Dimension | Measuring knowledge | Measuring Attitudes | Measuring Behaviours |
|---|---|---|---|---|
| HAIS-Q | Password Management | It's acceptable to use my social media passwords on my work accounts. (R) | It's safe to use the same password for social media and work accounts. (R) | I use a different password for my social media and work accounts. |
| HAIS-Q | Password Management | I am allowed to share my work passwords with colleagues. (R) | It's a bad idea to share my work passwords, even if a colleague asks for it. | I share my work passwords with colleagues. (R) |
| HAIS-Q | Password Management | A mixture of letters, numbers, and symbols is necessary for work passwords. | It's safe to have a work password with just letters. (R) | I use a combination of letters, numbers, and symbols in my work passwords. |
| HAIS-Q | Email Use | I am allowed to click on any links in emails from people I know. (R) | It's always safe to click on links in emails from people I know. (R) | I don't always click on links in emails just because they come from someone I know. |
| HAIS-Q | Email Use | I am not permitted to click on a link in an email from an unknown sender. | Nothing bad can happen if I click on a link in an email from an unknown sender. (R) | If an email from an unknown sender looks interesting, I click on a link within it. (R) |
| HAIS-Q | Email Use | I am allowed to open email attachments from unknown senders. (R) | It's risky to open an email attachment from an unknown sender. | I don't open email attachments if the sender is unknown to me. |
| HAIS-Q | Internet Use | I am allowed to download any files onto my work computer if they help me do my job. (R) | It can be risky to download files on my work computer. | I download any files onto my work computer that will help me get the job done. (R) |
| HAIS-Q | Internet Use | While I am at work, I shouldn't access certain websites. | Just because I can access a website at work, doesn't mean that it's safe. | When accessing the internet at work, I visit any website that I want. (R) |
| HAIS-Q | Internet Use | I am allowed to enter any information on any website if it helps me do my job. (R) | If it helps me do my job, it doesn't matter what information I put on a website. (R) | I assess the safety of websites before entering information. |
| HAIS-Q | Social Media Use | I must periodically review the privacy settings on my social media accounts. | It's a good idea to regularly review my social media privacy settings. | I don't regularly review my social media privacy settings (R). |
| HAIS-Q | Social Media Use | I can't be fired for something I post on social media. (R) | I feel comfortable posting on social media things I wouldn't normally say in a public setting. (R)* | I don't post anything on social media before considering any negative consequences. |
| HAIS-Q | Social Media Use | I can post what I want about work on social media. (R) | It's risky to post certain information about my work on social media. | I post whatever I want about my work on social media.(R) |
| HAIS-Q | Mobile Devices | When working in a public place, I have to keep my laptop with me at all times. | When working in a café, it's safe to leave my laptop unattended for a minute. (R) | When working in a public place, I leave my laptop unattended. (R) |
| HAIS-Q | Mobile Devices | I am allowed to send sensitive work files via a public Wi-Fi network. (R) | It's risky to send sensitive work files using a public Wi-Fi network. | I send sensitive work files using a public Wi-Fi network.(R) |
| HAIS-Q | Mobile Devices | When working on a sensitive document, I must ensure that strangers can't see my laptop screen. | It's risky to access sensitive work files on a laptop if strangers can see my screen. | I check that strangers can't see my laptop screen if I'm working on a sensitive document. |
| HAIS-Q | Incident Reporting | If I see someone acting suspiciously in my workplace, I should report it. | If I ignore someone acting suspiciously in my workplace, nothing bad can happen. (R) | If I saw someone acting suspiciously in my workplace, I would do something about it. |
| HAIS-Q | Incident Reporting | I must not ignore poor security behaviours by my colleagues. | Nothing bad can happen if I ignore poor security behaviours by a colleague. (R) | If I noticed my colleague ignoring security rules, I wouldn't take any action. (R) |
| HAIS-Q | Incident Reporting | It's optional to report security incidents. (R) | It's risky to ignore security incidents, even if I think they're not significant. | If I noticed a security incident, I would report it. |
| CS-S | Integrity | Storing data on digital platforms is not always secure. | | |
| CS-S | Integrity | Information and documents I have stored on digital platforms may be lost or deleted. | | |
| CS-S | Integrity | | Sharing data on digital platforms does not involve any risk. (R) | |
| CS-S | Integrity | It is possible for third parties to access information and documents stored on digital platforms. | | |
| Own | Governance | I know where to find DNV's cybersecurity policies (e.g, DMS-G), procedures, and guidelines when needed. | I am willing to adhere to DNV's cybersecurity policies (e.g. DMS-G), procedures, and guidelines. | I actively refer to and use DNV's cybersecurity policies (e.g. DMS-G) and procedures to guide my actions when dealing with potential security concerns. |
| Own | Priority | | | My business area has prioritized raising the cybersecurity awareness in the past year. |

*Rephrased from the original item: It doesn't matter if I post things on social media that I wouldn't normally say in public. (R)

Appendix 3. The interview questions

**General Questions**
1. When you hear "CS," what comes to mind?
2. What is your main working area? Office, customer related, field work
3. How long have you been working in the organisation? How would you describe the organisation's current state in terms of CS?
    - Do you think the organisation's current CS state should be better? How?
4. Do you feel that our team/line manager prioritizes cybersecurity and integrates it into our daily work practices? Why is that?
5. Do you feel comfortable and safe reporting cybersecurity incidents?
    - Do you know how to report CS incidents?

**Challenges and improvement**
6. In your view, what is preventing the organisation from becoming even better / achieving the ideal state in CS?
7. What challenges, if any, do you face when trying to prioritize CS in your role?
    - Can you think of any factors that can affect the degree to which you prioritize CS?
8. **According to you, what improvements would you suggest enhancing CS at DNV?**
    - How can we raise CS awareness in the organisation?
    - How should the CS improvement efforts be communicated to employees?

**Training and Governance**
9. How would you describe your experience with CS training at the organisation?
    - Do you feel that you have sufficient CS training? Why?
    - What aspects of the training have been most helpful?
    - Do you have any suggestions how to improve current CS related training?
    - How do you identify scams/phishing attempts <u>e.g. links</u>? And what do you do then?
10. How familiar are you with the organisation's CS procedures and policies?
    - How easy can you find them?
    - In what ways do you find them useful?
    - Are there any aspects you think could be improved?
11. Is there anything else you would like to add about CS that we have not covered in this interview?
    - Do you have any comments to the survey that you received.

Access Rights (optional)
12. How would you describe the role of access control in supporting CS at DNV?
13. What has been your experience with implementing or managing access control?
    - Are there any challenges you encounter?
    - To what extent do you feel access control is utilized by your coworkers?

Document Classification (optional)
14. How do you see document classification contributing to CS at the organisation?

15. How clear are the distinctions between labels like "unlabeled," "open," "restricted," "confidential," or "secret"?
    - Do you find the guidelines for using these labels easy to follow?
16. What has been your experience with classifying documents?
    - Do you think document classification is commonly practiced among your coworkers? Why or why not?

**Closing Questions**
17. Is there anything else you think we should discuss regarding CS that we haven't covered yet?

**Appendix 4. Survey and Interview responses on general CS improvement suggestions by country**

**Country 1's key improvement suggestions**

*Leadership commitment and leaders promoting strong CS culture*

Participants highlighted the importance of leadership acting as role models and proposed that line managers routinely lead discussions, such as in performance evaluation meetings, to reinforce individual responsibility in maintaining CS. In addition, participants recommended designating a dedicated CS Responsible in each region to serve as a clear point of contact and ensure local accountability.

*Foster organisational learning and awareness initiatives*

Participants emphasised the need to strengthen organisational learning by increasing transparency around CS risks and incidents. Suggestions included regularly sharing real-world and practical scenarios, both from within the organisation and from other companies, to build relevance and contextual understanding. Participants believed that discussing internal incidents and near misses would help employees understand the types of threats the organisation faced, why certain behaviours matter, and how risks manifested across different contexts. Participants also advocated for the use of case studies and lessons learned to raise awareness and bridge the gap between policy and practice.

Participants also highlighted the importance of continuously evaluating CSC through regular surveys and feedback loops. They suggested establishing reward and recognition programmes to encourage proactive CS behaviours and integrating CS discussions into team meetings to keep risks top of mind. Ongoing feedback on phishing test performance and regular updates on emerging threats, such as AI-based fraud and evolving phishing tactics, were also recommended. Some participants also called for guidance about safe use of personal devices, such as mobile phones used for work-related email access.

To engage employees, participants proposed a mix of formats, including webinars, quick-reference flyers, lessons learned, practical do's and don'ts, and interactive discussions as part of routine meetings.

To reach all parts of the organisation, it was suggested that CS moments be made mandatory during internal meetings and that communication occur in smaller teams and through top management channels. This approach was viewed as particularly important to ensure that employees with diverse backgrounds and roles, including external employees, develop a shared understanding of core CS expectations.

*Establish CS governance and measures and ensure consistent CS practices across the organisation*

Participants emphasised the need for standardised CS guidelines and playbooks covering incident response, disaster recovery, and best practices across the organisation. Regular CS assessments and audits were also suggested to proactively identify vulnerabilities.

To ensure consistent CS practices across the organisation, participants recommended fostering greater uniformity while acknowledging local differences. They proposed creating a global database of CS cases and examples to support shared learning and alignment. It was also emphasised that CS information should be easily accessible to all employees, including external employees to consistency in CS practices. Participants suggested providing the organisation's official laptops to all workers, including external employees, and implementing a Bring Your Own Device (BYOD) policy with periodic security scans to maintain a consistent and secure working environment across regions.

*Make trainings effective, continuous, interactive, targeted, tiered, and available for <u>all</u> employees*

Participants emphasised that all employees, permanent, temporary, external, should receive the same level of CS basic training at the minimum to ensure consistent awareness. They advocated for smaller, more frequent refresher e-learning sessions featuring practical examples, interactive content, gamified quizzes, and real-world case studies. Simulated cyber-attacks and hands-on response exercises were seen as valuable for improving incident handling. Participants also called for regular phishing tests with targeted follow-up, unit- and region-specific tailoring, and ongoing evaluation of training effectiveness. Periodic awareness programmes, webinars, and in-person workshops were recommended, along with training on emerging fraud techniques and how to handle them.

*Make reporting for non-phishing easy, encouraged, and guided*

Participants emphasised the need to simplify and promote reporting of all CS-related concerns, not just phishing. They called for clearer, coordinated guidelines on what to report, where to report it, and how, along with improved communication to ensure employees feel confident and encouraged to report suspicious activity, concerns, and near-misses.

*Strengthening CS technical measures*

Participants recommended strengthening CS technical defences by adopting the Principle of Least Privilege for access rights and supporting nested security models within Active Directory. They advocated for reducing the use of email links and shifting to secure communication channels like APIs and internal platforms. Suggestions included implementing biometric authentication to reduce password reliance, using multi-factor authentication, and enforcing strong password policies. Participants also emphasised the need for secured, standardized tools for external employees who might not have full access to internal platforms, secured network devices (e.g. dongles) for travel, and external audits of CS measures, including antivirus protection and device security. For strengthening access control, participants suggested developing smart systems that could automatically assess the sensitivity of documents and recommend appropriate classification labels, helping to enforce data protection practices more efficiently and consistently.

Participants emphasised the need for the organisation to be as innovative as cybercriminals by continuously advancing its knowledge and capabilities. They suggested for proactive screening of new services and technologies before implementation to identify potential risks early. Understanding how emerging technologies, such as Artificial Intelligence, might impact CS was seen as essential.

**Country 2's key improvement suggestions**

*Improve leaders' CS competence and make leaders to be on the same page to promote strong CS culture*

Senior leaders and managers were expected by participants as CS role models, highlighting the critical role of leadership in shaping a strong CSC. Participants also suggested to localize CS expertise and accountability by region. One key suggestion was to ensure that all senior managers receive consistent CS training to align their understanding and handling of CS incidents. The importance of appointing a dedicated CS responsible person in each region was emphasised to ensure accountability and localised oversight.

Participants also suggested to avoid using punitive approaches, proposing instead the use of supportive performance improvement plans for employees who repeatedly violate CS policies. In addition, there was a strong sentiment to put efforts promoting a mindset where CS is a shared responsibility across the organisation, not just the domain of CS experts.

*Foster organisational learning and awareness initiatives*

Participants recommended sharing real-world CS incidents and near misses by explaining what happened, how, and the potential impact to make risks tangible and relatable, raising awareness. Regular updates on CS threats, including stats such as phishing click rates, were suggested to keep employees informed. Town halls and department meetings should be used to highlight lessons learned, supported by ongoing campaigns, training, and phishing simulations. Participants also proposed using internal social medial platforms for interactive announcements, increasing policy visibility through targeted reminders, and integrating "CS moments" into regular meetings. Addressing conflicting priorities, such as workload and customer deadlines, and recognising employees who demonstrate strong CS practices were also seen as key to building a more engaged and aware culture.

*Establish clear and consistent CS governance*

Participants highlighted the need for stronger alignment in CS policies and practices across the organisation's global operations. Participants pointed out inconsistencies, for example, while Country 2 had a no-link policy in emails to prevent phishing incidents, employees still frequently received legitimate links from colleagues in other regions. To address this, participants suggested minimising the use of links in emails as a global organisation's policy and transitioning toward portal-based communication, such as directing users to centralised internal platforms to share critical information.

In addition, there were suggestions to implement periodic CS reminders on login screens or desktops to reinforce awareness in daily workflows. Participants also proposed standardising best practices and incident response playbooks across regions, developing clear CS emergency plans and communication strategies, and consolidating all CS-related information in a single, easily accessible location. Finally, periodic reviews of policies and procedures were recommended to ensure they remain relevant, up-to-date, and easily understood, such as eliminating outdated resources on intranet to avoid confusion.

*Make trainings tiered, targeted, realistic, interactive*

Participants emphasised the need for more interactive, engaging, and role-specific CS training or e-learning. Suggestions included increasing the frequency of mandatory refresher e-learning sessions and using or coupling up with hands-on formats like phishing simulations, real-world scenarios, and in-person workshops. Tiered training programmes were recommended to match different roles and expertise levels. Participants also suggested for more sophisticated, realistic phishing tests, with targeted follow-up training for those who failed. Providing individual feedback on performance and ensuring training remains practical, frequent, and relevant were seen as key to maintaining awareness and vigilance.

Participants highlighted the need to expand CS training content to cover a broader range of threats and practical guidance. In addition to phishing, they suggested for increased focus on risks such as unsafe USB use, public Wi-Fi vulnerabilities, and secure information sharing both internally and externally. Emphasising the real-world consequences of CS breaches was seen as essential to build urgency and relevance. More advanced, realistic phishing simulations were recommended to better reflect actual threat scenarios. Overall, participants requested more training on the variety of CS threats and clear guidance on how to respond effectively.

*Encourage and ensure employees know what, how, and where to report CS concerns, near-misses and incidents*

Participants emphasised the importance of fostering a proactive CSC by encouraging employees to report any suspicious activities or concerns. They also suggested providing individual feedback on phishing test performance to raise awareness and support continuous learning.

*Improve CS technical infrastructure, password management and access control*

Participants suggested for adopting secure, automated data-sharing methods, such as APIs instead of email attachments, and consolidating security tools to reduce complexity and improve user experience. They stressed the need for a safe, standardised platform for non-native apps and more intuitive digital design to prevent workarounds. Suggestions also focused on stronger password management, implementing the least privilege access controls, increasing acceptance of nested security models in Active Directory, and exploring biometric authentication for improved security and usability.